\documentstyle[prb,aps,twocolumn]{revtex}

\newcommand{\y}{\'{\i}}

\begin{document}
\include{epsf}

\title{Role of the intrinsic surface state in the decay of image 
states at a metal surface}

\author{J. Osma$^{1}$, I. Sarr\y a$^{2}$, E. V. Chulkov$^{1}$,
J. M. Pitarke$^{2}$, and P. M. Echenique$^{1,3}$}

\address{	
$^1$ Departamento de F\y sica de Materiales, Facultad de Ciencias  Qu\y micas,
Universidad del Pa\y s Vasco/Euskal Herriko Unibertsitatea, Aptdo. 1072,
20080, San Sebasti\' an, Basque Country, Spain\\
$^2$ Materia Kondentsatuaren Fisika Saila, Zientzi Fakultatea, 
Euskal Herriko Unibertsitatea, 644 Posta kutxatila,\\
48080 Bilbo, Basque Country, Spain\\
$^3$ Unidad Asociada al Instituto de Ciencia de Materiales, 
Consejo Superior de Investigaciones Cient\y ficas,\\
Cantoblanco, 28049 Madrid, Spain}
\date{\today}
\maketitle 
\begin{abstract}
The role of the intrinsic surface state ($n=0$) in the decay of the first
image state ($n=1$) at the (111) surface of copper is investigated. Inelastic
linewidths are evaluated from the knowledge of the
imaginary part of the electron self-energy, which we compute, within the GW
approximation of many-body theory, by going beyond a free-electron description
of the metal surface. Single-particle wave functions are obtained by solving the
Schr\"odinger equation with a realistic one-dimensional model potential, and
departure of the motion along the surface from free-electron behaviour is
considered through the introduction of the effective mass. The decay of the
first image state of Cu(111) into the intrinsic surface state is found to result
in a linewidth that represents a $40\%$ of the total linewidth. The dependence of
linewidths on the momentum of the image state parallel to the surface is also
investigated.
\end{abstract}
%
\section{INTRODUCTION}
The presence of an electron in front of a solid surface redistributes the
charge in the solid. As a consequence, an attractive potential is induced,
which far from the surface approaches the long-range classical image
potential ${V_{\rm im}(z)=-g_s/4z}$ ($z$ being the distance from the
surface, $\displaystyle{g_s=(\epsilon_s-1)/(\epsilon_s+1)}$, and $\epsilon_s$
the static bulk dielectric constant; for a metal, $g_s=1$). If the bulk band
structure projected onto the surface presents an energy gap near the vacuum
level, an electron located in front of the surface cannot propagate into the
solid. Therefore, the electron may be trapped in the vacuum well, and an
infinite series of Rydberg-like states appears, which converges, 
for zero parallel momentum, towards the vacuum energy. These so-called image
states\cite{Eche,Fauster95,Osgood97} are localized in the vacuum region of the
surface, the penetration of the first ($n=1$) image wave
function into the solid varying typically between $4\%$ and $22\%$ 
\cite{Chulkov99}. As a
result, image states are almost decoupled from bulk electron scattering, and
they are much longer lived than bulk excitations: The lifetime of bulk
electrons with energies of $4$ eV above the Fermi level is approximately
one order of magnitude smaller than that of the first image state.
Furthermore, the lifetime of higher order image states ($n\geq1$) has been
predicted\cite{Eche} to scale  asymptotically with $n^3$, which makes the series
to be resolvable.

During the last decade the linewidth of image states has been measured by
inverse photoemission\cite{Passek92,Passek95}, two-photon photoemission
\cite{Scho88,Schuppler92,Wang96}, and time-resolved
two-photon photoemission
\cite{Hertel96,Wolf96,Harris97,Wolf98,Shumay98}. Recently, time-resolved
two-photon photoemission has been used in combination with the coherent
excitation of several quantum states, and the lifetime of the first six image
states on the (100) surface of copper has been accurately
determined\cite{Hofer97}.

Theoretical calculations of the linewidth of image states were first reported in
Refs.\onlinecite{Eche85} and \onlinecite{deAnd87}, within a many-body
free-electron description of the metal surface and with the use of simplified
models to approximate both initial and final electronic states and, also, the
screened Coulomb interaction. Later on, the decay of the first image state
on the (111) surfaces of copper and nickel metals to the $n=0$ crystal-induced
surface state was calculated\cite{Gao}, in terms of Auger transitions, with the
use of a three band model to describe the surface band structure. In Ref. 
\onlinecite{Gao}, 
hydrogenic-like states with no penetration into the
solid were used to describe the image-state wave functions, a simplified
parametrised form was used for the surface-state wave functions, and screening
effects were neglected. Self-consistent calculations of the linewidths of image
states on copper surfaces have been reported recently\cite{Chulkov98}, and good
agreement with experimentally determined decay times has been found. In
Ref.\onlinecite{Chulkov98}, the linewidths of image states were computed, within
the GW approximation of many-body theory\cite{Hedin}, by going beyond a
free-electron description of the metal surface. Single-particle wave functions
were obtained by solving the Schr\"odinger equation with a realistic
one-dimensional model potential\cite{Chulkov97}, and the screened interaction
was evaluated in the random-phase approximation (RPA)\cite{rpa}.

In this paper we focus our attention on the role that the crystal-induced
surface state ($n=0$) plays in the relaxation of the first image state at
the (111) surface of copper, which we find to represent a $40\%$ of the total
linewidth. We present self-consistent calculations along  the lines of
Ref.\onlinecite{Chulkov98}, and we also consider simplified  models for both the
electronic wave functions and the screened Coulomb  interaction, showing that a
detailed description of these quantities is  of crucial importance in the
understanding of the origin and magnitude  of linewidths of image states. We
account for potential variation parallel to the surface through the introduction
of the effective mass, and we find that the linewidth of the first image state
of Cu(111) is $20\%$ smaller than in the case of free-electron behavior along
the surface. Finally, we investigate the dependence of interband\cite{notem1}
linewidths on the momentum of the image state parallel to the surface, ${\bf
k}_\parallel$. Our results indicate that for image state total energies lying
below the top of the gap, the linewidth of the first image state of Cu(111) is
increased with ${\bf k}_\parallel\neq 0$ up to a $20\%$.

\section{THEORY}

\label{theory}

We assume translational invariance in the plane parallel to the surface,
which is taken to be normal to the $z$-axis, and we evaluate the inelastic
linewidth of the image state $\phi_1(z)\;e^{{\rm i}{\bf k}_{\parallel}\cdot{\bf
r}_{\parallel}}$  with energy $E_1 = \epsilon_1 +{\bf
k}_{\parallel}^2/(2m_1)$\cite{note0} (we use atomic units throughout,
i.e., $e^2=\hbar=m_e=1$), as the projection of the imaginary part of the
electron self-energy, $\Sigma({\bf r},{\bf r}';E_1)$, over the state itself:
\begin{equation}
\label{gamma1}
\Gamma=-2\int dz dz' \phi^{*}_{1}(z) 
{\rm Im}\Sigma(z, z', {\bf k}_{\parallel};E_1) \phi_{1}(z'),
\end{equation}
where $\Sigma(z,z',{\bf k}_{\parallel};E_1)$ represents the two-dimensional
Fourier transform of $\Sigma({\bf r},{\bf r}';E_1)$. 

In the GW approximation\cite{Hedin}, the self-energy is obtained by just
keeping the first term of the expansion in the screened interaction ($W$). Then,
after replacing the Green function by the zero order approximation, one finds
\begin{eqnarray}
\label{sigma1}
{\rm Im}\Sigma(z,z',&{\bf k}_{\parallel}&;E_{1})=\sum_{E_F \leq E_f \leq
E_{1}}
\int 
\frac{d^2 {\bf q}_{\parallel}}{(2 \pi)^2} \phi^{\ast}_f(z') \nonumber \\ 
&\times&{\rm Im}W^{\rm ind}(z,z',{\bf q}_{\parallel};E_1-E_f)\phi_f(z), 
\end{eqnarray}
where the sum is extended over a complete set of final states  
$\phi_f(z) e^{i ({\bf k}_{\parallel}+{\bf q}_{\parallel})\cdot{\bf
r}_{\parallel}}$ with energies $E_f = \epsilon_f + 
({\bf k}_{\parallel}+{\bf q}_{\parallel})^2 /(2m_f)$. $m_f$ is the effective
mass, which accounts for the departure of the motion along the surface from
free-electron behavior, $E_F$ is the Fermi energy, and
$W^{\rm ind}(z,z',{\bf q}_{\parallel},E)$ represents the two-dimensional Fourier
transform of the induced part of the screened interaction.

In particular, if only transitions into the crystal-induced $n=0$ surface state
$\phi_0(z) e^{i ({\bf k}_{\parallel}+{\bf q}_{\parallel})\cdot{\bf
r}_{\parallel}}$ with energy $E_0=\epsilon_0 + 
({\bf k}_{\parallel}+{\bf q}_{\parallel})^2/(2m_0)$ are considered, one finds
\begin{eqnarray}
\label{sigmas}
{\rm Im}\Sigma_{\rm s}(z,z',&{\bf k}_{\parallel}&;E_{1})=\int 
\frac{d^2 {\bf q}_{\parallel}}{(2 \pi)^2} \phi^{\ast}_0(z') \nonumber \\ 
&\times&{\rm Im}W^{\rm ind}(z,z',{\bf q}_{\parallel};E_1-E_0)\phi_0(z), 
\end{eqnarray}
and introducing this contribution to ${\rm Im}\Sigma(z,z',{\bf
k}_{\parallel};E_{1})$ into Eq. (\ref{gamma1}) one finds:
\begin{eqnarray}
\label{surf}
\Gamma_{s}
=&&-2\int \frac{d^2 {\bf q}_{\parallel}}{(2 \pi)^2} \int dz dz' 
\phi^{*}_{1}(z)\phi^{*}_{0}(z')\nonumber \\
&&\times\:{\rm Im}W^{\rm ind}(z,z',{\bf q}_{\parallel};E_1-E_0)\phi_{0}(z)
\phi_{1}(z'). 
\end{eqnarray}

Three different models have been used for the evaluation of the screened
interaction, $W$. First, the specular-reflection model (SRM) of Ritchie and
Marusak\cite{Ritchie66} has been considered. In this model, bulk electrons are
assumed to be specularly reflected at the surface, the interference between the 
ingoing and outgoing waves being neglected, and the electronic charge
density abruptly terminates at the surface ($z=0$), which we
choose to be located half a lattice spacing beyond the last atomic layer. Within
this simplified model\cite{Eche95}, also called semiclassical infinite barrier
model (SCIBM), the screened interaction is obtained in terms of the wave-vector
and frequency dependent bulk dielectric function, which we evaluate in the RPA.
Secondly, for the vacuum contribution to the linewidth ($z>0$, $z'>0$) the surface
response function suggested by Persson and Zaremba\cite{Persson85} (PZ) has been used.
Finally, the screened interaction has been evaluated, as in Ref.\onlinecite{Chulkov98}, by
solving the RPA integral equation for the density response function of
inhomogeneous media in terms of the eigenfunctions of the one-electron effective
Hamiltonian. These eigenfunctions have been computed by solving the
Schr\"odinger equation with the realistic one-dimensional model potential
suggested in Ref.\onlinecite{Chulkov97}. This model potential uses as parameters the width 
and position of the energy gap at the $\bar{\Gamma}$ point (${\bf k_\parallel}=0$) 
and, also, the binding energies of both the $n=0$ crystal-induced surface state at
$\bar{\Gamma}$ and the first image state.

For the evaluation of $n=0$ and $n=1$ surface states, we have first used
simplified models for the wave functions inside and outside the solid. In the
vacuum side of the surface ($z>0$), $n=1$ and $n=0$ states have been
approximated by a parametrised $1s$-like hydrogenic wave function and a mere
exponential, respectively, which have been matched to a decaying wave
function in the crystal band-gap ($z<0$) obtained within a nearly free electron
two-band model\cite{deAnd87}:
\begin{equation}
\phi_{n}(z < 0) \sim e^{\Delta_{n} z} \cos{(G z + \delta_{n})}. 
\end{equation}
Here, $n=0$ and $n=1$ correspond to the crystal-induced surface state and the
first image state, respectively, $G$ is the limit of the Brillouin zone in the
direction normal to the surface, and
\begin{equation}
\Delta_{n} = \frac{1}{G} \sqrt{\frac{1}{4} E_{\rm gap}^2 - \bar{\epsilon}_n^2}, 
\end{equation}
where $E_{\rm gap}$ and $\bar{\epsilon}_{n}$ represent the energy gap and the
energy of the $n$ surface state with respect to the midgap, respectively. The
phase shift $\delta_n$ is given by
\begin{equation}
\delta_n = \frac{1}{2} \times \cases{
\pi - \;\tan^{-1}\left[\sqrt{\frac{1}{\eta^2_n}-1}\right],
& if $\;\;\;0 \leq \eta_n \leq 1$ \cr
\tan^{-1}\left[\sqrt{\frac{1}{\eta^2_n}-1}\right],
& if $-1 \leq \eta_n \leq 0,$ \cr}
\end{equation}
with $\eta_n=2\bar{\epsilon}_{n}/E_{\rm gap}$.

\begin{figure}[htb]
\centerline{
\epsfxsize = \columnwidth
\epsffile[116 352 469 691]{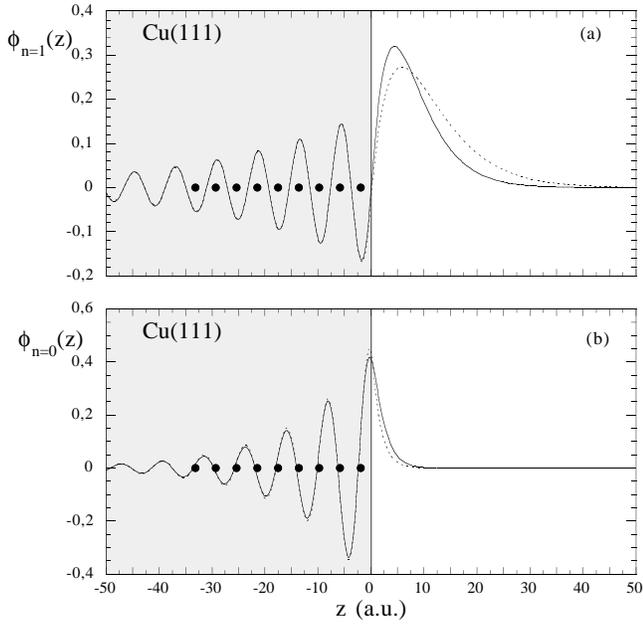}}
\begin{center}
\parbox{\columnwidth}{
\caption[]{\label{wave} Wave functions of both (a) $n=1$ image and (b) $n=0$
intrinsic surface states on Cu(111), as obtained within two different models: MWF
(solid line)  and AWF (dotted line) (see text). Full circles represent the
atomic  positions in the (111) direction. The geometrical electronic edge ($z = 0$) has
been chosen to be located half an interlayer spacing beyond the last atomic layer.
Notice the s-like and p-like characters of the image and intrinsic-surface state wave
functions, respectively.}}
\end{center}
\end{figure}

The image state on Cu(111) is
located right at the top of the gap ($\delta_1 \simeq 0.9\times\pi/2$), both the
hydrogenic-like wave function in the vacuum ($z>0$) and the decaying $s$-like
wave function in the bulk ($z<0$) having, therefore, nodes at the surface
($z=0$). The $n=0$ crystal-induced surface state on Cu(111) is located at the bottom of
the gap ($\delta_0 \simeq 0.2\times\pi/2$); thus, it is described by a $p$-like
wave function in the bulk. These approximate wave functions (AWF) are exhibited
in Fig. \ref{wave},  together with the corresponding wave functions that we
obtain by solving the Schr\"odinger equation with the one-dimensional potential
of Ref.\onlinecite{Chulkov97} (MWF). Both wave functions, AWF and MWF, coincide
within the bulk, but the hydrogenic-like wave function for the
$n=1$ image state appears to be less localized near the surface than our model
wave function. The $n=0$ and $n=1$ surface states on Cu(111) have binding
energies (measured with respect to the vacuum level) of $0.83$ and $5.32$ eV, 
respectively. The $n=1$ probability-density has a maximum at $4.3$ a.u. 
outside the crystal edge ($z=0$). The penetration into the bulk of $n=0$ 
and $n=1$ surface states is found to be, at the ${\bar \Gamma}$ point,
of $74.5\%$  and $22.1\%$, respectively.

\section{RESULTS AND DISCUSSION}	

\label{result}

The results of our calculations for the linewidth, $\Gamma_{\rm s}$, coming from
the decay of the $n=1$ image state into the $n=0$ intrinsic surface state on 
Cu(111) are presented in Table \ref{ttb1}, with the momentum of the image
electron parallel to the surface, ${\bf k}_\parallel$, set equal to zero. Here,
the linewidth has been split as follows:
\begin{equation} 
\label{split}
\Gamma_{\rm s} = \Gamma_{\rm vac} + \Gamma_{\rm sol} + \Gamma_{\rm inter}, 
\end{equation}
where $\Gamma_{\rm vac}$, $\Gamma_{\rm sol}$ and $\Gamma_{\rm inter}$ represent
vacuum, bulk and interference contributions, respectively, as obtained by
confining the integrals in Eq. (\ref{surf}) to either vacuum ($z>0,z'>0$), bulk
($z<0,z'<0$) or vacuum-bulk ($z\,_{<}^{>}0,z'\,_{>}^{<}0$) coordinates. First we
show our full RPA calculations, in which the screened interaction is obtained on the
basis of one-electron eigenfunctions computed from the realistic one-dimensional model
potential of Ref.\onlinecite{Chulkov97}. Within these calculations both
$n=0$ and $n=1$ surface-state wave functions are also obtained from the model potential
of Ref.\onlinecite{Chulkov97} (MWF), with either $m_0=1$
or $m_0=0.42$\cite{Goldmann85,Kevan83,Hulbert85}. Within the specular
reflection model\cite{note1} and the model suggested by Persson and
Zaremba\cite{note2} for the screened interaction, we have used the MWF $n=0$
and $n=1$ surface-state wave functions as well as the simplified models (AWF)
described in the previous section, with $m_0=1$.

Total linewidths, $\Gamma$, as obtained from the decay of the $n=1$ image
state on Cu(111) into any final state with energy $E_f$, $E_F\le E_f\le E_1$,
(see Eqs. (1) and (2)), are presented in Table II. Here, our full
RPA calculations are shown, with all wave functions computed from the
one-dimensional model potential of Ref.\onlinecite{Chulkov97} (MWF). Realistic
values of the effective mass of final states have been considered, according to
the experiment or to {\it ab initio} band structure calculations. As for the
$n=0$ surface state, we have used $m_0=0.42$, as in Table I, and for bulk states
we have chosen to increase the effective mass from our computed
value\cite{note4} of $m_f=0.22$ at the bottom of the gap to $m_f=1$ at the
bottom of the valence band.

Our full RPA calculations indicate that the decaying rate of the $n=1$ 
image state into the $n=0$ crystal-induced surface state results, for $m_0=1$,
in  a linewidth of $16$ meV, while use of the more realistic effective mass
$m_0=0.42$ leads to a linewidth of $12\,{\rm meV}$. With the use of either the
free-electron mass or more realistic effective masses for both bulk and
crystal-induced surface states,
$\Gamma_s$ approximately represents a $40\%$ of the total linewidth,
$\Gamma=37\,{\rm meV}$ ($m_f=1$) or $\Gamma=29\,{\rm meV}$ ($m_f\neq 1$). The
more realistic result of
$\Gamma=29\,{\rm meV}$ for the total linewidth is in good agreement with the
experimentally measured lifetime\cite{hbar} of $22\pm 3\,{\rm fs}$ at $25\,{\rm
K}$\cite{Wolf98,Shumay98}. Within the vacuum side  of the surface the
$n=1$ image state couples dominantly to the $n=0$ surface state (this coupling
approximately represents a $90\%$ of the total
$\Gamma_{\rm vac}$ linewidth); however, the coupling of image states with all
bulk crystal states occurring through the bulk penetration plays an important
role and cannot, therefore, be neglected if one is to accurately describe
the lifetime of image states. 

We note that simplified jellium models for the evaluation of the screened
interaction lead to unrealistic results for the contribution of the surface
state to the linewidth of image
states\cite{note3}. First, we compare our full RPA calculations (see Table I)
with the results we obtain, also with use of our model initial and final wave
functions (MWF), when our realistic screened interaction is replaced by that
obtained within the specular reflection model (SRM) and the model of Persson and
Zaremba (PZ). Bulk contributions to the linewidth are approximately well
described within the specular reflection model, small differences resulting from
an approximate description, within this model, of the so-called {\it begrenzung}
effects. As the approximate treatment of Ritchie and Marusak\cite{Ritchie66}
ignores the quantum mechanical details of the surface, this model fails to
describe both vacuum and interference contributions to the linewidth. These
quantum mechanical details of the surface are approximately taken into account
within the jellium model of Ref.\onlinecite{Persson85}, thus resulting in a
better approximation for the vacuum contribution to the linewidth. Discrepancies
between vacuum contributions obtained within this model (PZ) and our more
realistic full RPA calculations\cite{note3} appear as a result of the jellium
model of Ref.\onlinecite{Persson85} being accurate provided $q_\parallel/q_F$ 
and $\omega/E_F<<1$ ($q_F$ is the Fermi momentum, i.e., $E_F = q_F^2/2$).

In order to investigate the dependence of ${\Gamma}_{\rm s}$ 
on the details of both $n=1$ and $n=0$ wave functions, we
present in Table I calculations, within SRM and PZ models for the screened
interaction, in which our realistic wave functions are replaced by the
simplified models (AWF) described in the previous section. As the
hydrogenic-like wave function used to describe the $n=1$ image state on the
vacuum side of the surface presents an image-state charge gravity center localized
further away from the surface than our more realistic model wave function, both
vacuum and interference contributions to the linewidth are largely
underestimated within this approximation. Furthermore, we note that the
linewidth is highly sensitive to the details of the image-state wave functions. This
is a consequence of the critical behavior of the imaginary part of the non-local
self-energy coupling points near the surface, as we will discuss below.

\begin{figure}[htb]
\centerline{
\epsfxsize = \columnwidth
\epsffile[71 99 536 672]{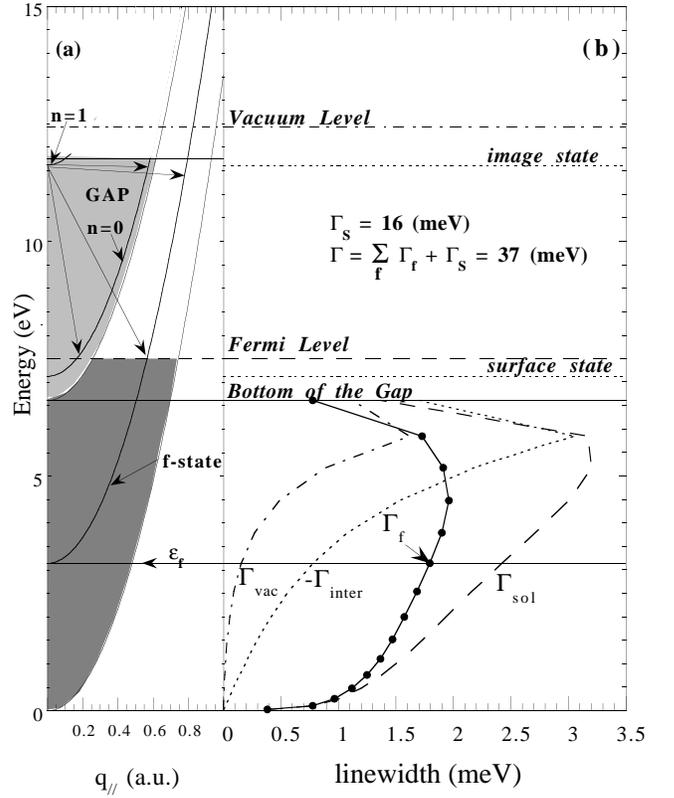}}
\begin{center}
\parbox{\columnwidth}{
\caption[]{\label{band} (a) Electronic surface band structure at the (111) surface
of copper. (b) Vacuum ($\Gamma_{\rm vac}$), bulk ($\Gamma_{\rm sol}$), and
interference  ($-\Gamma_{\rm inter}$) contributions to the linewidth
($\Gamma_f$) coming from the decay  into the various $f$ bulk crystal states.
The arrows determine the available phase space in the decay from the
$\bar{\Gamma}$ point of the $n=1$ image state into the unoccupied portion  of
both the $n=0$ surface state and the generic $f$ bulk state. Dispersion curves
of  these final states are depicted. The energy is measured with respect to the
bottom of the  valence band. All effective masses have been set equal to the
free-electron mass.}}
\end{center} 
\end{figure}

Now we focus on our full RPA calculation of the total linewidth of the $n=1$
image state on Cu(111) (see Table II), with all effective masses set equal to the
free-electron mass. We show in Fig.
\ref{band}b separate contributions  to the linewidth, $\Gamma$, coming from the
decay into the various $f$ bulk  crystal states, $\Gamma_f$, such that
\begin{equation}
\Gamma = \sum_f \Gamma_f + \Gamma_{\rm s}.
\end{equation}
Fig. \ref{band}a exhibits the bulk band structure projected onto the 
(111) surface of copper. The arrows indicate the available
phase space in the decay of the $n=1$ image state at the
$\bar\Gamma$ point (${\bf k}_\parallel$=0) into the unoccupied portion of
the $n=0$ surface state and a generic $f$ bulk state, which are
represented by their characteristic
$\epsilon_0 + q_\parallel^2/2$ and $\epsilon_f +
q_\parallel^2/2$ parabolic dispersions, respectively. $\Gamma_{\rm vac}$, 
$\Gamma_{\rm sol}$ and  $-\Gamma_{\rm inter}$ contributions to $\Gamma_f$ are
represented, together with the total contribution, $\Gamma_f$, as a
function of $\epsilon_f$.

 The lower edge, at the $\bar\Gamma$ point, of the energy gap projected onto 
the Cu(111) surface lies below the Fermi level ($E_g<E_F$) and, consequently,
the decay from the (${\bf k}_{\parallel} = 0$) image state occurs through 
finite parallel momentum transfer. Hence, as the coupling of the image state
with the crystal occurring through the tails of bulk states outside the
crystal is expected to be dominated by vertical transitions ($q_\parallel\simeq
0$), vacuum contributions to the $\Gamma_f$  linewidth are very small,
especially for those  bulk states located at the bottom of the valence band
(decay into these states is only allowed for large values of the momentum
transfer; also, their vacuum penetration is small). Actually, the coupling  of
the
$n=1$ image state with all bulk states taking place at the vacuum  side results
in a linewidth of only
$5$ meV, which approximately represents a $10\%$  of the total $\Gamma_{\rm vac}$
linewidth \cite{notefig2}.

A realistic description of motion along the surface can be approximated
by introduction of the effective mass, as described above. The effective mass
of all final states with energies $E_f$, $E_f\le E_f\le
E_1$, is found to be smaller than the free-electron mass, thus both $\Gamma_s$
and $\Gamma$ being about $20\%$ smaller than in the case of free-electron
behavior along the surface (see Tables I and II). This is the result of two
competing effects: First, there is the effect of the decrease of the available
phase space, which is easily found to scale as $\sqrt{m}$. Secondly, as the
effective mass decreases the decay from the image state occurs, for a given
energy transfer, through smaller parallel momentum transfer, which may result in
both enlarged and diminished screened interactions, depending on momentum and
energy transfers. This is illustrated in Fig. 3, where the impact of the
introduction of the effective mass on the evaluation of the various
contributions (vacuum, bulk, and interference) to both
$\Gamma_f$ and $\Gamma_s$ (see Eq. (9)) is exhibited through the percentage ratio
\begin{equation}
R_{\Gamma_f} =100\,\frac{
\Gamma_f(m_f\neq 1)}{\Gamma_f(m_f=1)}, 
\end{equation}
as a function of the final state energy $\epsilon_f$. If the screened
interaction were independent of the parallel momentum transfer, all ratios
would scale as $\sqrt{m}$, which is represented in Fig. 3 as $R_{\Delta
q_\parallel}=100\sqrt{m_f}$. Instead, as the parallel momentum transfer
decreases the screened interaction is predominantly larger, which results in
the ratio $R_{\Gamma_f}$ to be larger than $R_{\Delta q_\parallel}$,
especially in the case of vacuum contributions to the linewidth which are
expected to be dominated by vertical transitions. When the decay from the image
state may occur through very small parallel momentum transfer $q_\parallel$,
[this is the case of final states that are just below the bottom of the gap and
also the case of the $n=0$ surface state], a decrease in $q_\parallel$ may
result in a diminished screened interaction (see Fig. 4), thus
$R_{\Gamma_f}$ being slightly smaller than $R_{\Delta q_\parallel}$ for these
states.

Now, we analyze the behavior of the imaginary part of the image-electron
self-energy, which results in vacuum and interference contributions to the 
linewidth to be comparable in magnitude and opposite in sign (see Table
\ref{ttb1}). Fig.
\ref{self} shows full RPA (solid line) and SRM (dashed-dotted line) calculations
of
$-{\rm Im}\Sigma_{\rm s}(z,z', {\bf k}_{\parallel}=0, E_1)$ (see Eq.
(\ref{sigmas})) for Cu(111) with all effective masses set equal to the
free-electron mass, together with the $n=1$ image-state wave function, as a
function of the
$z'$-coordinate and for $z=7.4\,{\rm a.u.}$. We note that
the probability for  electron-hole pair creation (the dominant channel for the
decay of these states  is provided by this process) is underestimated within the
SRM.

The coupling between electronic states is well known to be maximum, within the
bulk, at the position of the electron. Nevertheless, as the electron moves into the
vacuum the maximum of the imaginary part of the electron self-energy stays (see
Fig. \ref{self}) near the surface ($z=0$), as demonstrated, within a jellium model of the
surface, by Deisz {\it et al}\,\cite{Deisz93}. Hence, for any given value of
$z>0$, main vacuum and interference contributions to the linewidth are
determined (see Eq. (\ref{gamma1})) by the specific shape of the image-state
wave function in regions $A$ and
$B$ of Fig. \ref{self}. As the image-state on Cu(111) is located right at
the top of the energy gap and the corresponding wave function has, therefore, a node at
the surface ($z\simeq 0$), an inspection of Fig. \ref{self} leads us to the
conclusion that vacuum and interference contributions to the linewidth are
comparable in magnitude and opposite in sign. On Cu(100) the image state is
located close to the center of the gap, and the corresponding wave function has
a node at $z\simeq 1.3$ a.u. On this surface, total vacuum and interference contributions
to the linewidth are  still opposite in sign, though interference contributions
coming from the decay into states at the bottom of the valence band are now positive
due to their minor vacuum penetration. If
the image state were located at the bottom of the gap, matching at the surface would
occur at maximum amplitude and the total interference contribution might be positive, as
the sign of the image-state wave function would be the same on both sides just around
the surface ($z=0$).

Finally, we investigate the dependence of interband\cite{notem1}
linewidths on the momentum of the image electron parallel to the surface, ${\bf
k}_\parallel$. First, we use the MWF image state wave function evaluated at the
${\bar \Gamma}$ point and introduce, within this model, the dependence on ${\bf
k}_\parallel$ of the quasiparticle self-energy, as indicated in Eq. (2).
We find that for image state total energies lying below the top of the gap
($k_\parallel\simeq 0.11\,{\rm a.u.}$), the linewidth of the first image
state of Cu(111) increases less than $4\%$. Secondly, we also account for the
change of the $z$-dependent initial wave function along
the dispersion curve of the image state, by solving the Schr\"odinger equation
for a one-dimensional model potential that we build following
Ref.\onlinecite{Chulkov97} with various values of $k_\parallel$: $0.06$,
$0.09$, and $0.10\,{\rm a.u.}$. The penetration into the bulk of the $n=1$ image
state with these values of $k_\parallel$ is found to vary from $22.1\%$ at
$k_{\|} = 0$ to
$22.6\%$, $24.1\%$ and $26.2\%$ at $k_{\|} = 0.06$ a.u., $0.09$ a.u. and
$0.10$ a.u., respectively. Our results, as obtained with all effective
masses set equal to the free-electron mass, are presented in Table III.
Though the penetration of the image state wave function increases with
$k_\parallel$, the amplitude of this wave function on the bulk side and near the
jellium edge decreases, thus the absolute value of both bulk and interference
contributions to the linewidth decreasing with $k_\parallel$.
Nevertheless, the total overlap between image state and final wave functions
becomes more efficient as $k_\parallel$ increases, which results in larger
values of the total linewidth. Also, it is interesting to notice that
especially sensitive to the variation of the momentum of the image
state parallel to the surface is the contribution to the linewidth from damping
into the $n=0$ surface state.

\section{SUMMARY}

\label{summary}

We have reported calculations of the inelastic broadening of the first image
state at the (111) surface of copper, and we have investigated, in particular,
the role that the intrinsic crystal-induced surface state plays in the decay of
this image state. We have presented self-consistent RPA calculations, by going
beyond a free-electron description of the metal surface. We have also
considered simplified models for both the electron wave functions and the
screened Coulomb interaction, showing that a detailed description of these
quantities is of crucial importance in the understanding of the origin of
linewidths of image states. We have accounted for potential variation parallel
to the surface through the introduction of the effective mass.

We have analyzed the origin and magnitude of the various contributions to the
linewidth. Though the dominant contribution to the decay of the first image
state into the crystal-induced surface state comes from the
coupling between image and surface states within the vacuum part of the
surface, it appears to be approximately canceled out by the contribution
from the interference between bulk and vacuum coordinates. For the vacuum
contribution to the decaying rate into the intrinsic surface state, we have
found that it approximately represents a $90\%$ of the total vacuum contribution.
We also conclude that the coupling of image states with all bulk crystal states
occurring through the bulk penetration plays an important role in the
determination of lifetimes, and that this penetration cannot be
neglected if one is to accurately describe the lifetime of image states.

We have found, within our full RPA scheme, that in the case of ${\bf
k}_\parallel=0$ the decaying rate of the first image state on Cu(111) into the
intrinsic surface state results, with all effective masses set equal to the
free-electron mass, in a linewidth of $16$ meV, while use of more realistic
effective masses leads to a linewidth of $12\,{\rm meV}$. With the use of
either the free-electron mass or more realistic effective masses for both bulk
and crystal-induced surface states, $\Gamma_s$ approximately represents a $40\%$
of the total linewidth, $\Gamma=37\,{\rm meV}$ ($m_f=1$) or $\Gamma=29\,{\rm
meV}$ ($m_f\neq 1$). The more realistic result of $\Gamma=29\,{\rm meV}$ for the
total linewidth is in good agreement with recent experimental results
reported in Ref.\onlinecite{Wolf98}.

We have investigated the dependence of interband linewidths on the momentum
of the image electron parallel to the surface, showing that for image state
total energies lying below the top of the gap the linewidth of the first image
state increases with $k_\parallel$ up to a $20\%$. We conclude that this
increase appears mainly as a consequence of the change of the $z$-dependent
initial wave function with $k_\parallel$, and our results indicate that
the contribution to the linewidth from damping into the $n=0$ surface state is
responsible for the dependence of the total linewidth with the momentum
parallel to the surface.

\section{ACKNOWLEDGEMENTS}

The authors gratefully acknowledge A. Rubio, E. Zarate and M. A. Cazalilla for
fruitful discussions in connection with this research. This project has
been supported by Eusko Jaurlaritza (Basque Country), the Ministerio de Educaci\'
on y Cultura (Spain), and Iberdrola S. A..

\newpage
\begin{table}[h]
\caption[]{Calculated linewidth (in meV) coming from the decay of the $n=1$ image
state on Cu(111) into the $n=0$ intrinsic surface state, as obtained within
three different models for the description of the surface response and two different
models for the description of both initial and final wave functions (see text). RPA
accounts for our full RPA realistic calculation of the screened interaction, SRM for the
simplified specular reflection model of Ritchie and Marusak \cite{Ritchie66}, and PZ for
the vacuum side surface response suggested by Persson and Zaremba \cite{Persson85}. MWF
accounts for the wave functions obtained by solving the Schr\"odinger equation with the
realistic one-dimensional model potential of  Ref.\onlinecite{Chulkov97}, and AWF for the
approximate model described in the text. The effective mass of the $n=1$ image
state has been set equal to the free-electron mass, and for the $n=0$ surface
state we have used either $m_0=1$ or $m_0=0.42$. The momentum of the image
electron parallel to the surface is set equal to zero.}
\begin{tabular}{lcrrrrc} 
Surf. Res. & Wave function & $\Gamma_{{\rm vac}}$ & $\Gamma_{{\rm sol}}$ & 
$\Gamma_{{\rm inter}}$ & $\Gamma_{s}$ \\
\hline
RPA ($m_0=1$) & MWF & 42 & 16 & -42 & 16 \\
RPA ($m_0\neq 1$)& MWF & 29 &  8 & -25 & 12 \\
\hline
SRM ($m_0=1$)& MWF & 11 & 12 & -17 & 6 \\
SRM ($m_0=1$)& AWF & 2 & 15 & -9 & 8 \\
\hline
PZ ($m_0=1$) & MWF & 55 & - & - &   -  \\
PZ ($m_0=1$) & AWF & 12 & - & - &   -  \\
\end{tabular}
\label{ttb1}
\end{table}

\begin{table}[h]
\caption[]{Calculated total linewidth (in meV) coming from the decay of the $n=1$ image
state on Cu(111) into any unoccupied final state with $E_f < E_1$, as obtained
within our full RPA scheme with all wave functions computed from the
one-dimensional model potential of Ref.\onlinecite{Chulkov97}. As in Table I,
the effective mass and the momentum parallel to the surface of the $n=1$ image
state have been set equal to the free-electron mass and equal to zero,
respectively. As for final (bulk and intrinsic surface) states, we have used
either $m_f=1$ or the realistic effective masses ($m_f=1$) described in the
text. Contributions to the linewidth from decay into bulk states,
$\Gamma-\Gamma_s$, are displayed in parentheses.}
\begin{tabular}{ cccccc } 
Surf. Res. & $\Gamma_{{\rm vac}}$ & $\Gamma_{{\rm sol}}$ & 
$\Gamma_{{\rm inter}}$ & $\Gamma$ \\ 
\hline
RPA($m_f=1$) & 47(5) & 44(28) & -54(-12) & 37(21) \\
RPA($m_f\neq 1$) & 34(5) & 32(24) & -37(-12) & 29(17) \\
\end{tabular}
\label{ttb2}
\end{table}
\newpage
\begin{table}[h]
\caption[]{Calculated total linewidth (in meV) of the first image state on
Cu(111), computed within our full RPA scheme with all wave functions computed
from the one-dimensional model potential of Ref.\onlinecite{Chulkov97}, as a
function of the momentum of the image electron parallel to the surface,
$k_\parallel$ (see text). All effective masses have been set equal to the
free-electron mass. Contributions to the linewidth from decay into the $n=0$
crystal-induced surface state, $\Gamma_s$, are displayed in parenthesis.}
\begin{tabular}{ cccccc } 
$k_{\|}$ & $\Gamma_{{\rm vac}}$ & $\Gamma_{{\rm sol}}$ & 
$\Gamma_{{\rm inter}}$ & $\Gamma$ \\ 
\hline
0.0000 & 47 (42) & 44 (16) & -54 (-42) & 37 (16) \\
0.0570 & 48 (44) & 40 (13) & -49 (-38) & 39 (19) \\
0.0912 & 50 (45) & 32 (8)  & -38 (-29) & 44 (24) \\
0.1026 & 50 (44) & 28 (6)  & -31 (-23) & 47 (27) \\
\end{tabular}
\label{ttb3}
\end{table}

\begin{figure}[htb]
\centerline{
\epsfxsize = \columnwidth
\epsffile[58 122 497 677]{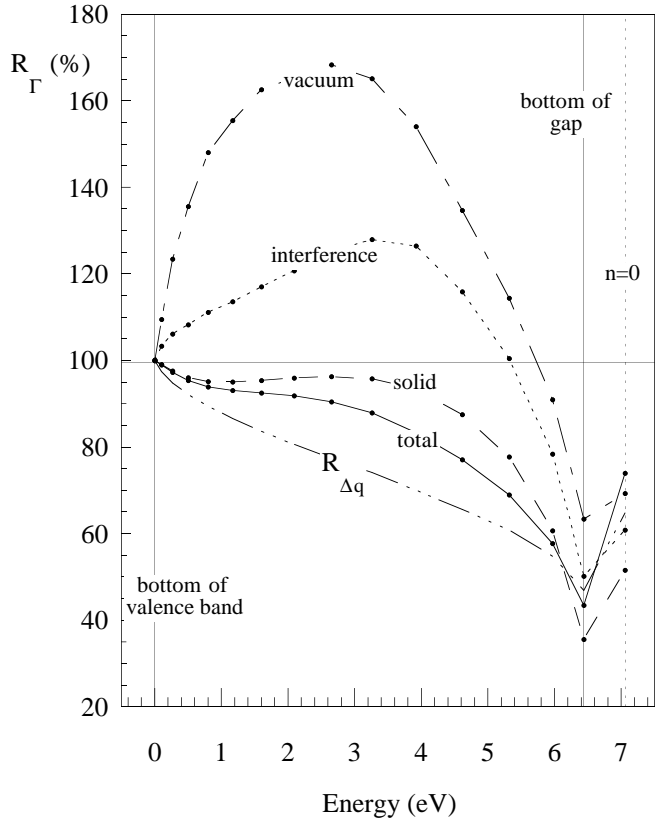}}
\begin{center}
\parbox{\columnwidth}{
\caption[]{\label{percentage} Ratio $R_{\Gamma_f}$ of Eq. (10), as a function of
the final state energy. The effective mass and the momentum parallel to the
surface of the $n=1$ image state have been set equal to the free-electron mass
and equal to zero, respectively. As for final (bulk and intrinsic
surface) states, we have used the realistic effective masses ($m_f \neq 1$) described
in the text. Here, $R_{\Delta q_\parallel}=100\sqrt{m_f}$.}}
\end{center} 
\end{figure}
\begin{figure}[htb]
\centerline{
\epsfxsize = \columnwidth
\epsffile[67 122 497 677]{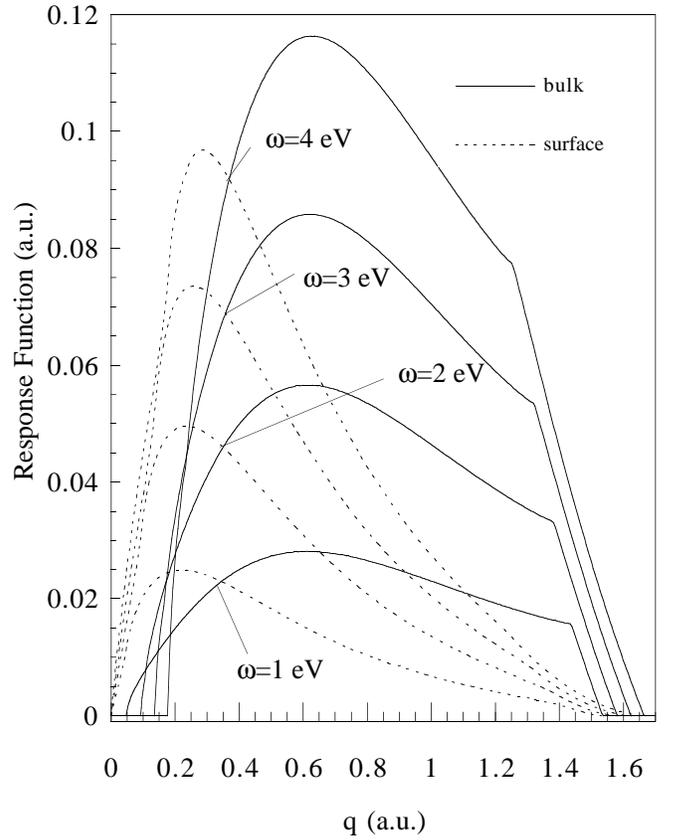}}
\begin{center}
\parbox{\columnwidth}{
\caption[]{\label{response} Imaginary part of bulk (solid lines) and surface
(dotted lines) response functions, ${\rm
Im}\left[-1/\epsilon(q,\omega)\right]$ and
${\rm Im}\left[-g_s(q_{\|},\omega)\right]$, which describe the screened interaction
far from the surface into the bulk and into the vacuum, respectively. Bulk
response functions have been evaluated in the RPA, whereas surface response
functions have been evaluated within the specular reflection model of the
surface (SRM)\cite{Eche95} with the RPA for the bulk dielectric function 
\cite{rpa}.}}
\end{center} 
\end{figure}
\begin{figure}[htb]
\centerline{
\epsfxsize = \columnwidth
\epsffile[73 116 512 675]{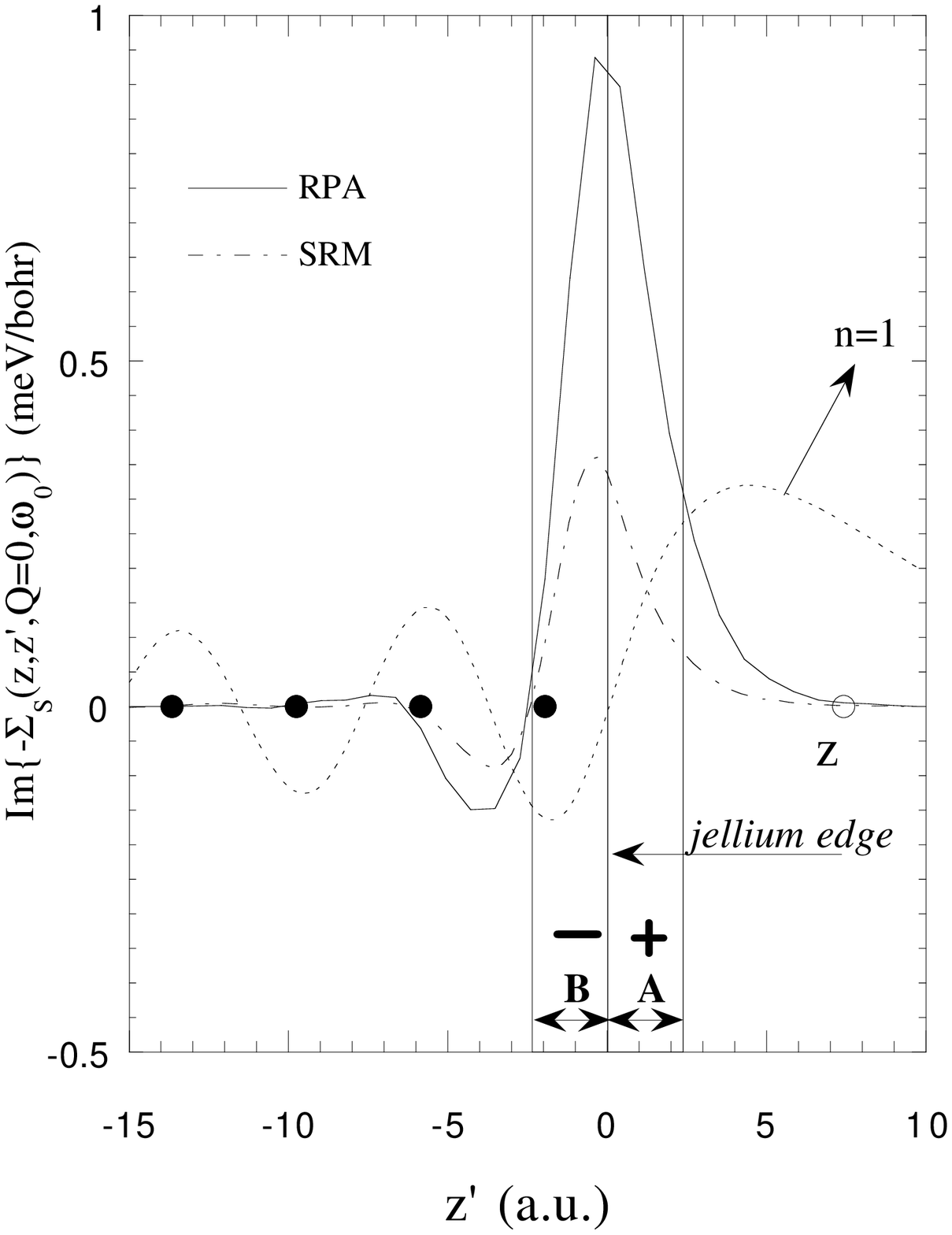}}
\begin{center}
\parbox{\columnwidth}{
\caption[]{\label{self} Full RPA (solid line) and SRM (dashed-dotted line) calculations
of $-{\rm Im}\Sigma_{\rm s}(z,z', {\bf k}_{\parallel}=0, E_1)$, as obtained from
Eq. (\ref{sigmas}) and as a function of the
$z'$ coordinate. The dotted line represents the $n=1$ image state
wave function (MWF). The value of $z$ ($z=7.4$ a.u.) is indicated by an open circle. The
sign appearing in regions A and B accounts for the sign of the product between the
imaginary part of the self-energy and the image state wave function. Full circles 
represent the atomic positions in the (111) direction. The geometrical electronic
edge ($z = 0$) has been chosen to be located half an interlayer spacing beyond the last
atomic layer. All effective masses have been set equal to the free-electron
mass.}}
\end{center} 
\end{figure}


\begin{thebibliography}{20}
\bibitem{notem1} In these processes the image electron decays into a state having
a different $z$-dependent (the plane of the surface is assumed to be
perpendicular to the $z$-axis) wave function.
\bibitem{note0} The effective mass of the image state is taken to be equal to the
free-electron mass, i.e., $m_1=1$ (see, e.g., Ref.\onlinecite{Fauster95}).
\bibitem{note1} The electron density parameter has been taken to be $r_{\rm
s}=2.58$, so as to reproduce the Fermi energy of $7.53$ eV obtained within our
model calculation of the ground state.
\bibitem{note2}
The parameters entering the surface response function of Persson and
Zaremba\cite{Persson85} have been taken to be $m_{opt}=1.68$, $h=1.84$ and
$\xi=0.49$.
\bibitem{hbar}
The lifetime, $\tau$, is related to the linewidth, $\Gamma$, of Eqs.
(\ref{gamma1}) and (\ref{surf}) by $\Gamma\cdot\tau=\hbar=660$ meV $\cdot$ fs.
\bibitem{note3} We find that the impact of the band structure
in the evaluation of the screened interaction is not large, and differences
between our full RPA calculations, on the one hand, and the SRM and PZ
results, on the other hand, appear as a consequence of simplifications
introduced, within a jellium model of the surface, in both SRM and PZ
approaches.
\bibitem{notefig2}
Notice that both $\Gamma_{vac}$ and $\Gamma_{inter}$ contributions 
to the linewidth present
a maximum for energies just below the bottom of the energy gap. This 
is due to the fact that the vacuum penetration of the bulk states increases
as one goes up in energy and then decreases just below the gap due to 
the stronger bulk character of these states reproducing the bottom of the gap. 
This maximum appears here as a cusp because of the relatively thin  
slab used in the present calculation. This sharp feature will be significantly
smoother for thicker films.
\end{thebibliography}
\end{document}